\documentclass[epj,nopacs]{svjour}
\usepackage{graphicx}

\title{
{\Large
Resonance Form-Factors:
$L_8$ determination  at
Next-to-Leading Order  in $\rm 1/N_C$ } }

\author{J.J. Sanz-Cillero}

\institute{
        Department of Physics, Peking University,
        Beijing 100871, P. R. China   \\
\email {cillero@th.phy.pku.edu.cn}
   }

\begin{document}

\newcommand{\xs}{\mbox{$x(\sigma)$}}
\newcommand{\vs}{\vbox{\vskip 1cm plus .3cm minus .3cm}}
\def\question#1{{{\marginpar{\small \sc #1}}}}
\newcommand{\bean}{\begin{eqnarray*}}
\newcommand{\eean}{\end{eqnarray*}}
\newcommand{\gapproxeq}{\lower
.7ex\hbox{$\;\stackrel{\textstyle >}{\sim}\;$}}
\newcommand{\lapproxeq}{\lower
.7ex\hbox{$\;\stackrel{\textstyle <}{\sim}\;$}}
\newcommand{\arr}{\mbox{$\rightarrow$} }
\newcommand{\Apr}{\mbox{\overline{ \rm p} } }
\newcommand\lsim{\mathrel{\rlap{\lower4pt\hbox{\hskip1pt$\sim$}}
    \raise1pt\hbox{$<$}}}
\newcommand\gsim{\mathrel{\rlap{\lower4pt\hbox{\hskip1pt$\sim$}}
    \raise1pt\hbox{$>$}}}
\newcommand{\ba}{\begin{array}}
\newcommand{\ea}{\end{array}}
\newcommand{\nn}{\nonumber}
\newcommand{\mathbold}{\bf}
\newcommand{\be}{\begin{equation}}
\newcommand{\ee}{\end{equation}}
\newcommand{\bear}{\begin{eqnarray}}
\newcommand{\eear}{\end{eqnarray}}
\newcommand{\tab}{\hspace*{0.5cm}}
\newcommand{\cen}{\hspace*{7.0cm}}
\newcommand{\ii}{\'{\i}}
\newcommand{\II}{\'{\I}}
\newcommand{\sla}{\hspace*{-0.2cm}\slash  }
\newcommand{\slag}{\hspace*{-0.25cm} \slash}

\newcommand{\rvac}{\,|0\rangle}
\newcommand{\lvac}{\langle 0|\,}
\newcommand{\ket}{\,\rangle}
\newcommand{\bra}{\langle \,}
\newcommand{\eqn}[1]{(\ref{#1})}
\newcommand{\cO}{{\cal O}}
\newcommand{\bel}[1]{\be\label{#1}}
\newcommand{\mL}{\mathcal{L}}
\newcommand{\mA}{\mathcal{A}}
\newcommand{\mB}{\mathcal{B}}
\newcommand{\mC}{\mathcal{C}}
\newcommand{\mD}{\mathcal{D}}
\newcommand{\mM}{\mathcal{M}}
\newcommand{\mO}{\mathcal{O}}
\newcommand{\mF}{\mathcal{F}}
\newcommand{\mT}{\mathcal{T}}
\newcommand{\mR}{\mathcal{R}}
\newcommand{\mG}{\mathcal{G}}
\newcommand{\mP}{\mathcal{P}}
\newcommand{\Frac}[2]{\frac{\displaystyle #1}{\displaystyle #2}}
\newcommand{\Int}{\displaystyle{\int}}

\abstract{ This talk presents a systematic procedure for the
computation of the SS-PP correlator beyond the large--$N_C$ limit.
The present calculation is carried on within a perturbative $1/N_C$
framework . By constraining the meson form-factors at leading order
in $1/N_C$, one obtains a one-loop spectral function well behaved at
short distances. The Weinberg sum-rules get modified, gaining an
extra contribution suppressed by $1/N_C$. This leads to a prediction
for the low energy chiral perturbation theory coupling $L_8^r(\mu)$
at the one-loop level, i.e., up to next-to-leading order in $1/N_C$.
} \PACS{ {CODIGO1}{CODIGO2} }

\maketitle

\section{The large--$N_C$ limit}

\vspace*{-0.2cm} \tab Quantum Chromodynamics  \ (QCD) has provided a
wide understanding of hadronic processes. At long distances, the
theory becomes non-perturbative.  An approach that has been found to
be useful  is to consider QCD in the limit of an infinite number of
colors $N_C\to\infty$, keeping $N_C \alpha_s$ finite~\cite{NC}.
Assuming confinement, large--$N_C$ QCD results equivalent to a
theory with an infinite number of narrow-width mesons, with the
matrix elements given by the tree-level amplitudes. Hadronic loops
are suppressed by $1/N_C$.

A crucial \ ingredient  \ in light-meson interactions  \ is  \
chiral symmetry.  \ The QCD lagrangian with $n_f$~massless flavors
is invariant under the chiral group $SU(n_f)_L\otimes SU(n_f)_R$,
which gets spontaneously broken into the vector subgroup
$SU(n_f)_{L+R}$~\cite{chpt}. Actually, in the large--$N_C$ limit the
symmetry group gets enlarged into $U(n_f)_L\otimes U(n_f)_R$ and it
is spontaneously broken into $U(n_f)_{L+R}$, producing $n_f^2$
Nambu-Goldstone bosons~\cite{anomaly}. Below the $\rho(770)$ vector
resonance multiplet, the Goldstones are the only hadrons in the
spectrum. Their interaction can be described through an effective
field theory based on chiral symmetry, namely chiral perturbation
theory ($\chi$PT)~\cite{chpt,U3chpt}.  Symmetry imposes stringent
constraints on the structure of the low energy interaction, so it is
an essential ingredient to recover the proper low-energy QCD
behavior. Hence, the interactions between mesons (Goldstones and
resonance) must be given by a chiral theory of resonances
(R$\chi$T)~\cite{therole,biblia}.

Following Refs.~\cite{letter,Cata,quantumloops,functional}, this
talk proposes a systematic program to aboard  the analysis of
hadronic amplitudes beyond the leading order in $1/N_C$ (LO).  The
example of the isovector SS-PP correlator is studied:
\begin{equation}
\Pi(t)=i\Int dx^4 e^{iqx}\bra T\{J(x) J(0)^\dagger -
J_5(x)J_5(0)^\dagger \} \ket \, ,
\end{equation}
with $J=\bar{d}u$ and $J_5=i\bar{d}\gamma_5 u$, and $t=q^2$.
The calculation is carried on within the chiral limit.

\vspace*{-0.2cm}
\section{A program for calculations at NLO in $1/N_C$}

\subsection{R$\chi$T lagrangian}

\tab Although the large--$N_C$ spectrum contains an infinite number
of hadrons, the Green-functions that are chiral order parameters are
mainly governed the lightest states. In general, one truncates the
tower of states and works under a minimal hadronical approximation
(MHA)~\cite{MHA}, keeping the minimal number of resonance multiplets
enough to fulfill the short-distance constraints. Though the
truncation of the spectrum  induces
uncertainties~\cite{correlator,L8largeNC}, these can be estimated
through the last absorptive channel included the
calculation~\cite{letter}.

The particles included in our meson lagrangian are the chiral
Goldstones and the lightest $1^{--},\,1^{++}, \, 0^{++},\, 0^{--}$
resonance multiplets. Since we work within the large $N_C$
framework, the hadrons are classified into $U(n_f)$ multiplets.

The hadronic lagrangian contains all the available operators
consistent with chiral symmetry. Our building blocks are the
resonance fields and the chiral tensors containing the
Goldstones~\cite{therole,biblia}. For the spin--1 fields we use the
antisymmetric tensor formalism~\cite{therole,spin1fields}. In
addition, in order to avoid a wrong growing behavior of the
Green-functions at high energies, the operators only contain tensors
up to $\cO(p^2)$. The operators of the lagrangian can be organized
on the number of resonance fields:
\begin{equation}
\mL_{\rm R\chi T }   = \mL_{\rm \chi PT}^{(2)} + \sum_{\rm R_1}
\mL_{\rm R_1} + \hspace*{-0.2cm}  \sum_{\rm R1,R_2} \hspace*{-0.2cm}  \mL_{\rm R_1,R_2} +
\hspace*{-0.2cm}  \sum_{\rm
R1,R2,R3} \hspace*{-0.2cm}  \mL_{\rm R_1,R_2,R_3}+ ...
\end{equation}
where the first term on the r.h.s. is the $\cO(p^2)$  $\chi$PT
lagrangian~\cite{chpt} and the second term, linear in the resonance
fields, was long ago constructed in Ref.~\cite{therole}. For the
form-factors we are interested on (those with two mesons in the
final state), only the operators with three or less resonance fields
are relevant. They extra  terms contain new couplings $\lambda_{\rm
R_1R_2}, \, \lambda_{\rm R_1R_2R_3}$~\cite{biblia,letter}.

\subsection{Form-factors at large--$N_C$}

\tab In the large $N_C$ limit the correlator is given by the
tree-level exchange of Goldstones and resonances.  At the one
loop level one may find the two-meson absorptive diagrams shown in
Fig.~\eqn{fig.diagrams}. The calculation is carried on
within perturbation theory
and no Dyson-Schwinger resummation is performed. Hence, all the lines
in Fig.~\eqn{fig.diagrams} stand for tree-level meson propagators.
\begin{center}
\begin{figure}
\center{\includegraphics[angle=-90,width=8cm]{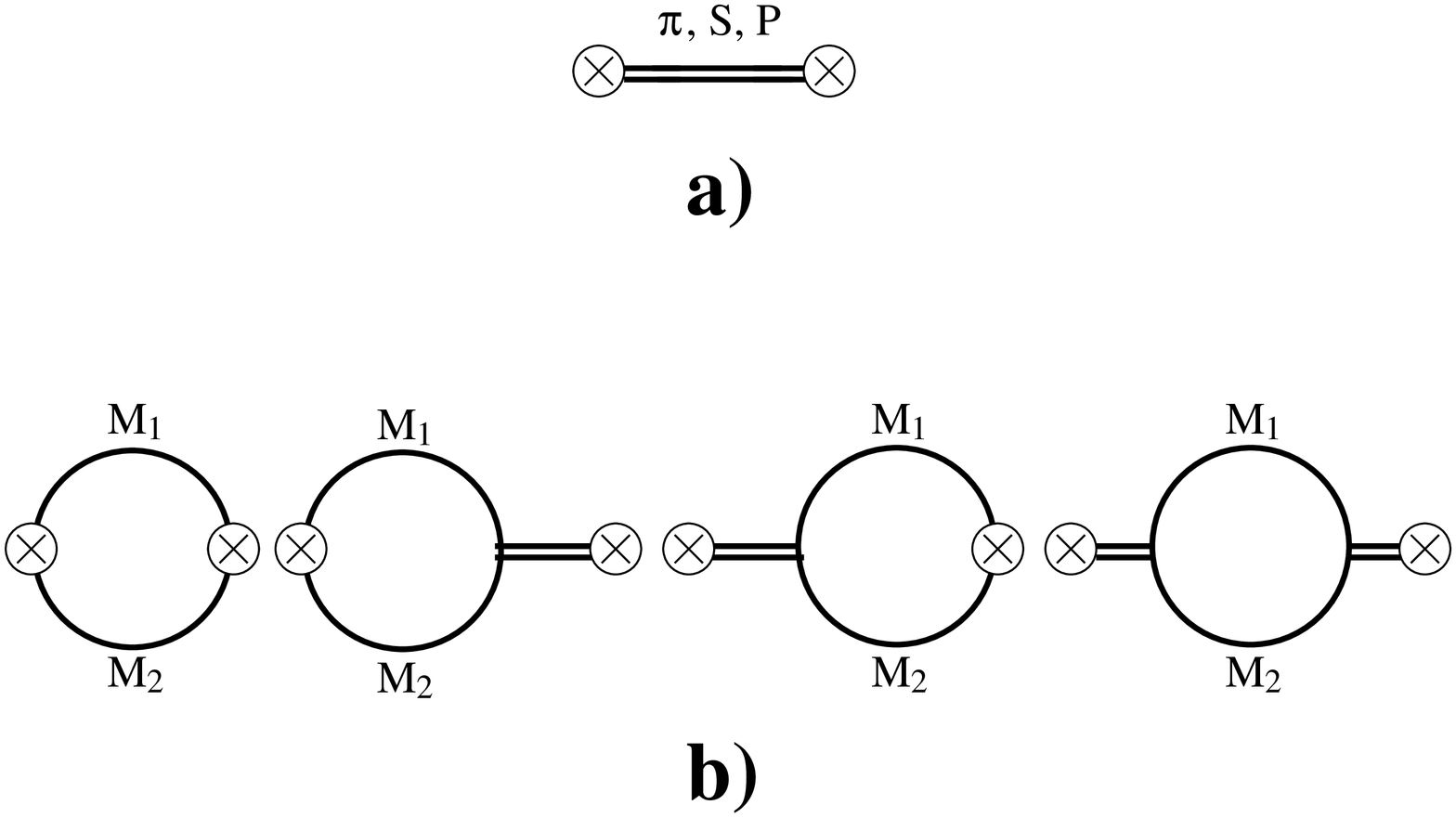}}
\vspace*{-0.5cm}
\caption{\small{Absorptive contributions for the SS-PP correlator:
Tree-level meson exchange (a), and one-loop diagrams (b). }}
\label{fig.diagrams}
\end{figure}
\end{center}

At LO, the only absorptive cuts are one-particle cuts. At NLO, one
may also have intermediate two-meson states. For a particular
two-particle cut $\rm M_1M_2$, its contribution to the spectral
function is in general proportional to some form factor squared:
\begin{equation}
\mbox{Im}\Pi(t)_{_{\rm M_1M_2}} \propto
\left|\mF_{_{\rm M_1M_2}}(t)\right|^2\, .
\end{equation}

By means of  quark-counting rule arguments~\cite{Brodsky}, it is
usually accepted that the pion scalar form factor vanishes at
infinite momentum. It  accepts an unsubtracted dispersion relation
which leads to the usual monopolar expression
$\mF_{\pi\pi}(t)=\frac{M_S^2}{M_S^2-t}$ and to a contribution
Im$\Pi(t)_{\pi\pi}$ to the spectral function which vanishes at
$t\to\infty$..

Demanding \ each  \ separate \ absorptive contribution
Im$\Pi(t)_{_{\rm M_1M_2}}$  to vanish at least as fast as
Im$\Pi(t)_{\pi\pi}$  leads to a series of constraints for the
form-factors at LO in $1/N_C$. Furthermore, the $\pi\pi$ and $R\pi$
form factors become determined in terms of the resonance
masses~\cite{letter}.

\subsection{Derivation of $\Pi(t)$}

\tab It is known \ from the  \  operator \  product \  expansion
(OPE)~\cite{ope} that the SS-PP correlator accepts an unsubtracted
dispersion relation:
\begin{equation}
\Pi(t)\, =\, \frac{1}{2\pi i}\, \oint \, dt'\, \frac{\Pi(t')}{t'-t}
\, =\, \frac{1}{\pi}\int_{0}^\infty \, dt' \,
\frac{\mbox{Im}\Pi(t')}{t'-t}\, . \label{eq.disp}
\end{equation}

At LO, the spectral function is given by a sum of delta functions
centered on the meson masses. Within the MHA, this gives the
large--$N_C$ correlator
\begin{equation} \Pi(t)\, =\, 2 B_0^2\,
\left[ \frac{8 c_m^2}{M_S^2-t} - \frac{8 d_m^2}{M_P^2-t}+
\frac{F^2}{t} \right] \, .
\end{equation}

At NLO, the spectral function contains as well finite two-particle
contributions Im$\Pi(t)_{_{\rm M_1M_2}}$, related to the two-meson
form-factors $\mF_{_{\rm M_1M_2}}(t)$. By means of
Eq.~\eqn{eq.disp}, one finds that the two-meson cuts contribute to
the correlator with a finite part, $\Delta \Pi(t)_{_{\rm M_1M_2}}$,
and a NLO renormalization of the scalar and pseudo-scalar masses and
couplings. Hence, the whole correlator up to NLO shows the general
structure:
\begin{eqnarray}
\Pi(t) = 2 B_0^2 \left[ \Frac{8 c_m^{r\, 2}}{M_S^{r\, 2}-t} -\Frac{8
d_m^{r\, 2}}{M_P^{r\, 2}-t} +\Frac{F^2}{t}\right]
\nn \\
\qquad \qquad \qquad \qquad \qquad + \displaystyle{\sum_{\rm
M_1M_2}}\Delta \Pi(t)_{_{\rm M_1M_2}}\, ,
\end{eqnarray}
with the finite contribution from the ${\rm M_1M_2}$ cut,
\begin{eqnarray}
\Delta\Pi(t)_{_{\rm M_1M_2}}&=&\lim_{\epsilon\to 0^+}\left[
  \Int_{\mR_\epsilon}
  \,  \frac{dt}{\pi}\frac{\mbox{Im}\Pi(t')_{_{\rm M_1M_2}}}{t'-t}
\right.
\\
&&\left. -\frac{2}{\pi\epsilon} \lim_{t'\to
M_R^2}\left(\frac{(M_R^2-t')^2\, \mbox{Im}\Pi(t')_{_{\rm
M_1M_2}}}{t'-t}\right)\right]\, ,  \nn
\end{eqnarray}
with $R=S,P$ the corresponding $s$--channel resonance, and the
interval $\mR_\epsilon=[0,M_R^{r\, 2}-\epsilon] \cup [M_R^{r\,
2}+\epsilon, +\infty)$
%
%
%
%

It is interesting to recall that no new couplings are required after
demanding an unsubtracted dispersion relation for $\Pi(t)$. To fix
the correlator at NLO, one just needs to specify the value of the
renormalized masses $M_S^r,\, M_P^r$ and couplings $c_m^r,\, d_m^r$.

\subsection{Matching OPE up to NLO in $1/N_C$}

\tab  In the high energy limit, the two--meson contribution is found
to behave as
\begin{equation}
\Delta \Pi(t) =  \frac{F^2}{t}\, \delta_{\rm NLO}^{(1)} \,+ \,
\frac{F^2 M_S^2}{t^2}\, \left(\delta_{\rm NLO}^{(2)} +
\widetilde{\delta}_{\rm NLO}^{(2)}\ln{\frac{-t}{M_S^2}}\right)\, ,
\end{equation}
where the NLO constants $\delta_{\rm NLO }^{(1)},\, \delta_{\rm
NLO}^{(2)}$ and $\widetilde{\delta}_{\rm NLO}^{(2)}$ depend on the
decay constant $F$ and the resonance masses $M_R$~\cite{letter}.

The one-loop R$\chi$T correlator can be now matched to OPE in the
deep euclidian region, finding similar expressions to the Weinberg
sum-rules, but now containing extra terms,  NLO in $1/N_C$:
\begin{eqnarray}
- 8 c_m^{r\, 2}  + 8 d_m^{\, 2} + F^2\, (1+ \delta_{\rm NLO}^{(1)})
&=& 0 \, ,
 \\
- 8 c_m^{r\, 2} M_S^{r\, 2} + 8 d_m^{\, 2} M_P^{r\, 2}+ F^2 M_S^2\,
\delta_{\rm NLO}^{(2)} &\simeq& 0 \, ,  \label{eq.D4const}
\end{eqnarray}
where the dimension--4 OPE condensate is much smaller than each
single term in Eq.~\eqn{eq.D4const} and it can be safely
neglected~\cite{PI:02}. The matching is fulfilled by demanding that
the $\frac{1}{t^2}ln{\frac{-t}{M_S^2}}$ term also vanishes, this is,
$\widetilde{\delta}_{\rm NLO}^{(2)}=0$.

These relations allow fixing the resonance couplings up to NLO:
\begin{eqnarray}
c_m^{r\, 2} &=& \frac{F^2}{8}\Frac{M_P^{r\, 2}}{M_P^{r\, 2}-M_S^{r\,
2}} \left[ 1 + \delta_{(1)}-\frac{M_S^2}{M_P^2}\delta_{(2)} \right]
\, ,
\\d_m^{r\, 2}&=& \frac{F^2}{8}\Frac{M_S^{r\, 2}}{M_P^{r\,
2}-M_S^{r\, 2}}  \left[ 1 + \delta_{(1)}- \delta_{(2)} \right]\, .
\end{eqnarray}

 When considering just $\pi\pi$ and $R\pi$ cuts, one finds that
after imposing the QCD short distance conditions everything becomes
determined in terms of the renormalized masses $M_R^r$.

At low energies, the contribution from higher and higher thresholds
becomes more and more suppressed. The two-resonance cuts are
neglected in the present work. The uncertainty from the truncation
is estimated from the $P\pi$ contribution, the higher threshold
under consideration.

\subsection{Recovery of $\chi$PT at low energies}

\tab One of the main advantages of working within a chiral invariant
framework is the recovery of $\chi$PT at low energies even at the
loop level. The one--loop R$\chi$T calculation exactly reproduces
the one--loop $\chi$PT expression.  This provides a prediction for
the value of the renormalized low energy constant (LEC),
$L_8^r(\mu)$, in terms of R$\chi$T parameters. The two expressions
match at any $\mu$ and R$\chi$T generates the exact $L_8^r(\mu)$
running found in $\chi$PT~\cite{chpt}. There is not a specific
saturation scale but a relation between renormalized LECs and
renormalized R$\chi$T parameters.

In the low energy limit, the R$\chi$T expression can be expanded in
powers of $t$:
\begin{equation}
\Pi(t)= B_0^2\left\{\frac{2 F^2}{t}  +  32 \bar{L}_8^{U(3)} +
\frac{3}{16\pi^2}\left(1-\ln{\frac{-t}{M_S^2}}\right)
\hspace*{-0.15cm}+\cO(t)\right\}\, ,
\end{equation}
with  the constant
\begin{eqnarray} \bar{L}_8^{U(3)} \quad &=& \quad
\frac{F^2}{16}\, \left[\frac{1}{M_S^{r\, 2}}+\frac{1}{M_P^{r\,
2}}\right]
\\
&& \times \, \left\{ 1\,+ \, \delta_{\rm NLO}^{(1)}\, - \,
\frac{M_S^{r\, 2}}{M_S^{r\, 2}+M_P^{r\, 2}}\delta_{\rm
NLO}^{(2)}\right\}
 - \frac{3\, \Delta}{256\pi^2}\, . \nn
\end{eqnarray}
The $\cO(1)$ constant $\Delta$ is given in Ref.~\cite{letter}. It
comes from the two-particle contribution $\Delta \Pi(t)_{_{\rm
M_1M_2}}$ and is a function of $F$ and $M_R$.


Comparing this result with $U(3)-\chi$PT~\cite{U3chpt}, one gets a
prediction for the renormalized LEC $L_8^r(\mu)$:
\begin{equation}
L_8^{r}(\mu)_{_{U(3)}}= \bar{L}_8^{U(3)} -
\frac{3}{512\pi^2}\ln{\frac{\mu^2}{M_S^2}}\, .
\end{equation}

The last step is to integrate out the chiral singlet $\eta_0$.  In
the large--$N_C$ limit, the $\eta_0$ is the ninth Goldstone and it
is massless~\cite{anomaly}. However, it gains mass due to  higher
order corrections. Naively, one would expect that the effect of the
$\eta_0$ mass in the one-loop diagrams
would be next-to-next-to-leading order, this is, suppressed by
$\frac{1}{N_C^2}$. Actually, since we study an energy limit below
the $\eta_0$ threshold ($t\ll M_0^2$), the first effect from  the
$\eta_0$ mass appears  at order $\frac{1}{N_C}\ln{\frac{1}{N_C}}$.
Thus, the $SU(3)-\chi$PT constant is finally related to the $U(3)$
prediction through~\cite{U3chpt,letter}
\begin{equation}
L_8^r(\mu)_{_{SU(3)}} \quad = \quad L_8^r(\mu)_{_{U(3)}} -
\frac{1}{384\pi^2}\ln{\frac{M_{\eta_0}^2}{\mu^2}}\, .
\end{equation}

\vspace*{-0.4cm}
\section{Conclusions}

\tab One of the main advantages of a chirally invariant theory of
resonances is that the symmetry properties ensures the right
recovery of the QCD low energy limit, $\chi$PT, even at the loop
level.

The absorptive $\chi$PT  logarithms are exactly reproduced by our
result at long distances. This removes the large--$N_C$ ambiguity
about the renormalization scale of saturation of $L_8^r(\mu)$. The
renormalized chiral coupling is given in terms of the renormalized
resonance effective parameters $c_m^r,\, d_m^r, \, M_S^r,\, M_P^r$.

The systematic $1/N_C$ expansion within the R$\chi$T framework
allows to derive Weinberg sum-rules beyond the leading order. This
fixes the value of the renormalized scalar and pseudo-scalar
couplings in terms of the renormalized resonance masses.

\acknowledgement{ Talk given at QNP'06, Madrid, Spain. Further
details can be found in Ref.~\cite{letter}. Work done in
collaboration with I.~Rosell and A.Pich.  It has been partially
supported by EU~RTN~Contract~CT2002-0311 and  China National Natural
Science Foundation under grant number 10575002 and  10421503. I want
to acknowledge the organizers of the conference for their work and
all the attentions received. }

\vspace*{-0.5cm}

\end{document}